\documentclass[%
 reprint,
 superscriptaddress,
preprintnumbers,
 amsmath,
 amssymb,
 aps,
 prc,
]{revtex4-1}

\usepackage{graphicx}
\usepackage{dcolumn}
\usepackage{bm}
\usepackage{multirow}
\usepackage{longtable}
\usepackage{tabularx}
\usepackage{url}
\usepackage[mathlines]{lineno}

\begin{document}

\preprint{APS/123-QED}

\title{Neutron capture cross section measurement of $^{238}$U at the n\_TOF CERN facility with C$_6$D$_6$ scintillation detectors in the energy region from 1 eV to 700 keV}

\author{F.~Mingrone}\affiliation{European Organization for Nuclear Research (CERN), Geneva, Switzerland}%
\affiliation{Dipartimento di Fisica e Astronomia, Universit\`{a} di Bologna, Italy}%
\affiliation{Istituto Nazionale di Fisica Nucleare, Bologna, Italy}%
\author{C.~Massimi}\affiliation{Dipartimento di Fisica e Astronomia, Universit\`{a} di Bologna, Italy}%
\affiliation{Istituto Nazionale di Fisica Nucleare, Bologna, Italy}%
\author{G.~Vannini}\affiliation{Dipartimento di Fisica e Astronomia, Universit\`{a} di Bologna, Italy}%
\affiliation{Istituto Nazionale di Fisica Nucleare, Bologna, Italy}%
\author{N.~Colonna}\affiliation{Istituto Nazionale di Fisica Nucleare, Bari, Italy}%
\author{F.~Gunsing}\affiliation{Commissariat \`{a} l'\'{E}nergie Atomique (CEA) Saclay - Irfu, Gif-sur-Yvette, France}%
\author{P.~\v{Z}ugec}\affiliation{Department of Physics, Faculty of Science, University of Zagreb, Croatia}%
\author{S.~Altstadt}\affiliation{Johann-Wolfgang-Goethe Universit\"{a}t, Frankfurt, Germany}%
\author{J.~Andrzejewski}\affiliation{Uniwersytet \L\'{o}dzki, Lodz, Poland}%
\author{L.~Audouin}\affiliation{Centre National de la Recherche Scientifique/IN2P3 - IPN, Orsay, France}%
\author{M.~Barbagallo}\affiliation{Istituto Nazionale di Fisica Nucleare, Bari, Italy}%
% ...
%

\collaboration{The n\_TOF Collaboration (www.cern.ch/ntof)}  \noaffiliation

\date{\today}

\begin{abstract}
The aim of this work is to provide a precise and accurate measurement of the $^{238}$U(n,$\gamma$) reaction cross section in the energy region from 1 eV to 700 keV. This reaction is of fundamental importance for the design calculations of nuclear reactors, governing the behaviour of the reactor core. In particular, fast reactors, which are experiencing a growing interest for their ability to burn radioactive waste, operate in the high energy region of the neutron spectrum. In this energy region most recent evaluations disagree due to inconsistencies in the existing measurements of up to 15\%. In addition, the assessment of nuclear data uncertainty performed for innovative reactor systems shows that the uncertainty in the radiative capture cross-section of $^{238}$U should be further reduced to 1-3\% in the energy region from 20 eV to 25 keV. To this purpose, addressed by the Nuclear Energy Agency as a priority nuclear data need, complementary experiments, one at the GELINA and two at the n\_TOF facility, were proposed and carried out within the 7th Framework Project ANDES of the European Commission. 

The results of one of these $^{238}$U(n,$\gamma$) measurements performed at the n\_TOF CERN facility are presented in this work. The gamma-ray cascade following the radiative neutron capture has been detected exploiting a setup of two C$_6$D$_6$ liquid scintillators. Resonance parameters obtained from this work are on average in excellent agreement with the ones reported in evaluated libraries. In the unresolved resonance region, this work yields a cross section in agreement with evaluated libraries up to 80 keV, while for higher energies our results are significantly higher.

\begin{description}
\item[PACS numbers] 25.40.Lw, 27.90.+b
\end{description}
\end{abstract}

\maketitle


\section{\label{sec:intro}Introduction}
In modern society energy has become one among the main fuels for social and economic development. Nowadays, the amount of energy used is massive, and it is destined to rise as population and wealth increase. Needless to say, energy-related activities have significant environmental impacts. The main and most addressed issue is the emission of greenhouse-gas that follows the production of energy through fossil fuels as coal, oil and natural gas, and which results in serious damage to climate, biodiversity and human health~\cite{iea_GHG_emission}. To avoid the business-as-usual dependence on fossil fuels, a map of future energy mix that incorporates alternative sources is needed. Several low-carbon resources should be part of this sustainable energy portfolio, and among them nuclear energy seems to be one of the few options available at scale to reduce carbon-dioxide emissions while providing a baseload generation~\cite{iea_WEO14}.

In this scenario, the development of improved reactor technologies that will lead to an intrinsically safe nuclear energy production is regarded as a necessary step to ensure a long term sustainable energy supply~\cite{IAEA13}. Towards this direction, the Generation IV International Forum (GIF)~\cite{GIF} works since 2001 to identify and select six nuclear energy systems that will be commercially available by 2030-2040. In addition, the concept of a nuclear subcritical device called accelerator driven system (ADS) is being developed both for energy production and radioactive waste disposal~\cite{ADS}.

The R\&D of these new reactor concepts require a level of accuracy and precision that challenges the present knowledge of nuclear data. The prediction of the behaviour of the reactor cores strongly depends on neutron-induced reaction cross-section data. In this context, the measurement of the $^{238}$U radiative capture cross section is of high priority and is part of the NEA High Priority Request List~\cite{nea_HPRL}, a compilation of the most relevant nuclear data requirements maintained by the Nuclear Energy Agency of the Organisation for Economic Co-operation and Development (OECD). In fact, $^{238}$U constitutes more than 90\% of nuclear fuel in power reactors, being one of the most important isotopes for neutron transport calculations in the active zone. In particular, for ordinary thermal nuclear reactors the uncertainties on fundamental design parameters such as the multiplication factor, the power peak, and the reactivity coefficient significantly depend on the $^{238}$U(n,$\gamma$) reaction cross-section for incident neutron energies from 0.0253 eV to 25 keV. Furthermore, the $^{238}$U capture cross section uncertainty affects the uncertainty of the Pu isotope density at the end of the fuel cycle (1.1\% for $^{239}$Pu, 0.2\% for $^{240}$Pu and 0.1\% for $^{241}$Pu~\cite{WPEC_subGroup}). 

The importance of the radiative capture reaction on $^{238}$U initiated many measurements of this cross section, mainly at the Oak Ridge Electron Linear Accelerator (ORELA)~\cite{Orela} and at the Geel Linear Accelerator (GELINA)~\cite{Gelina} facilities. In particular, an evaluation has been made by Derrien \textit{et al.} in 2005 including the most significative previous capture and transmission measurements~\cite{Derrien1, Derrien2}. Nevertheless, in the fast region of the neutron spectrum, which is fundamental for fast neutron reactor calculations, inconsistencies between published experimental data can reach 15\%, and the most recent evaluations disagree with each other by more than the uncertainties of the standards evaluation by Carlson \textit{et al.} \cite{Carlson}. In addition, a recent project has started, the Collaborative International Evaluated Library Organization (CIELO)~\cite{cielo}, which aims at producing an evaluated library common to Europe, North America and Asia. The focus of this project will initially be on a small number of the highest-priority isotopes, including $^{238}$U due to its key relevance within the nuclear industry.

To solve the present inconsistencies and to lower the uncertainty in the cross section down to 1-3\% in the energy range from 20 eV to 25 keV, complementary experiments, one at the EC-JRC laboratory GELINA and two at the CERN n\_TOF facility, were scheduled within task 1.2 of the FP7 project ANDES of the European Commission~\cite{andes_deliv}. The proposed objective was to combine results from different detection systems and independent experimental principles, obtained from measurements performed independently in different facilities, so as to minimize the overall systematic uncertainty.

This work presents the determination of high-accuracy radiative capture cross-section of $^{238}$U, starting from the measurement performed at the n\_TOF facility with an array of two C$_6$D$_6$ liquid scintillators, which has been optimized with respect to the background induced by sample-scattered neutrons, and takes advantage of the total energy detection technique. This experimental setup, together with the extremely high instantaneous neutron flux of the n\_TOF facility, opens the possibility of measuring the $^{238}$U capture cross-section also for neutron energies above 100 keV, an energy range of extreme interest for new nuclear technologies. In this work the $^{238}$U(n,$\gamma$) cross section has been analyzed from 1 eV up to 700 keV. 

\section{\label{sec:exp}Experiment}
The measurement was performed at the neutron Time-Of-Flight (TOF) facility of CERN, n\_TOF, which became operative in 2001 and has since been at the cutting edge of neutron cross section measurements. The pulsed neutron beam at n\_TOF is produced by spallation of 20 GeV/c protons from the CERN Proton Synchrotron accelerator on a water-cooled Pb target. A complete description of the n\_TOF facility can be found in Refs.~\cite{Gunsing_report, Gue_report}. Since the spallation mechanism is a remarkably powerful source of neutrons,  the instantaneous intensity of the n\_TOF neutron source of $\sim 2 \times10^{15}$ neutrons/pulse, is one of the highest worldwide. 

The pulsed neutron source is used together with a moderation system, so that the n\_TOF neutron beam covers about eleven orders of magnitude in energy from thermal to GeV. In 2012, when the $^{238}$U(n,$\gamma$) measurement was performed, borated water was used as moderator to minimize the production of 2.2-MeV $\gamma$ rays originating from neutron capture in water, which constitutes the main source of background in measurements of capture cross-sections in the keV neutron-energy region.

The facility has been designed on the basis of Monte Carlo FLUKA~\cite{Fluka_2014, Fluka_2005} simulations, and in particular the neutron source has been fully characterized together with its related $\gamma$ background. 

The measurement takes advantage of the first beam line operative at n\_TOF, which is about 200 m long and is designed to clean as much as possible the neutron beam from secondary particles profusely produced in spallation reactions. To this purpose, a sweeping magnet is placed at 145 m from the spallation target to deflect the remaining charged particles in the beam, and the beam tube is embedded in massive iron and concrete shielding to stop particles around the beam pipe. Two collimators 72 and 175 m downstream of the spallation target are used for shaping the neutron beam. For capture measurements the inner diameter of the second collimator is reduced to 1.8 cm to allow for a close geometry between sample and detectors. The first experimental area (EAR1), where the samples and detection systems are placed, is a 7.9 m long room starting at 182.3 m downstream of the spallation target. 

The neutron beam of the n\_TOF facility is constantly monitored by two independent low-background devices, without significantly affecting the neutron beam: the Silicon Monitor (SiMON)~\cite{simon}, with 4 silicon detectors looking at a 300 $\mu$m thick $^6$Li foil in the beam, and two MicroMegas detectors~\cite{Pancin_mgas, Belloni_mgas}, with $^{235}$U and $^{10}$B as neutron converter. The neutron flux of the n\_TOF facility for the 2009, 2010 and 2011 campaigns has been evaluated combining five independent measurements performed with different detectors, so as to cover the entire energy range with the highest precision reachable. A detailed description of the method can be found in Ref.~\cite{flux}. The flux related to this measurement, performed in 2012, was not included in the evaluation procedure, but it has been monitored at the experimental area with SiMON, showing that the shape of the flux has remained constant within 2\% (i.e within uncertainties) between 0.1 eV and 150 keV. In addition, the proton current, used to normalize the total neutron output, has been monitored with the pick-up detection system of the CERN Proton Synchrotron.

\begin{table*}
\caption{\label{tab:samples}Characteristics of the samples.}
\begin{ruledtabular}
\begin{tabular}{cccc}
      Sample & Size  & Mass  & Atomic density \\
      & (mm) & (g) & (atoms/barn) \\
      \hline
	$^{238}$U  &  $53.825 \times 30.12$ & $6.125 \pm 0.006$ & $(9.56 \pm 0.04)10^{-4}$ \\
	$^{nat}$Pb  & $53.77 \times 30.19$ & $9.44$ & $1.725 \times 10^{-3}$ \\
	$^{197}$Au & $53.30 \times 29.65$ & $9.213$ & $1.773 \times 10^{-3}$ \\
	$^{197}$Au & $53.30 \times 29.65$ & $1.547$ & $2.9 \times 10^{-4}$ \\
	$^{nat}$C & $53.35 \times 30.20$ & $28.89$ & $8.94 \times 10^{-2}$ \\
	$^{nat}$C & $53.35 \times 30.20$ & $14.638$ & $4.49 \times 10^{-2}$ \\
\end{tabular}
\end{ruledtabular}
\end{table*}

For capture measurements at n\_TOF, two different detection systems are available: a 4$\pi$ BaF$_2$ Total Absorption Calorimeter (TAC) ~\cite{Gue_TAC} and an array of deuterated benzene liquid scintillator detectors (C$_6$D$_6$ detectors). The measurement of this work has been carried out with two C$_6$D$_6$ scintillators placed face to face at 90$^\circ$ with respect to the beam, 9 mm upstream of the sample: one commercial Bicron and one custom-made developed at Forschungszentrum Karlsruhe (FZK)~\cite{Plag_FZK}. Both detectors are optimized to have a very low sensitivity to background signals induced by scattered neutrons. To this purpose the amount of detector material has been minimized, and only materials with low neutron capture cross-section. have been used For instance, the window of the photomultiplier is a custom Boron-free quartz window, and the housing of the FZK detector is entirely made of carbon fiber. 

The total energy detection technique has been exploited, combining the detection system described above with the Pulse Height Weighting Technique (PHWT)~\cite{Borella_C6D6} in order to assure the proportionality between the detection efficiency and the total energy released in the capture event. 

The $^{238}$U sample, which has been provided by the EC-JRC laboratory \cite{Gelina}, consists of an extremely pure metal plate, $6.125 \pm 0.006$ g in mass, containing less than 1 ppm of $^{234}$U, about 11 ppm of $^{235}$U and less than 1 ppm of $^{236}$U. The sample is approximately rectangular in shape, with an area of $1621.2 \pm 0.1$ mm$^2$, and covers about 97\% of the neutron beam. The effective area has been determined by an optical surface inspection with a microscope-based measurement system from Mitutoyo~\cite{Mitutoyo}.
To act in accordance with CERN radio protection regulations, the sample has been encased in $\sim 60$ $\mu$m and $\sim 75$ $\mu$m thick Al and Kapton foils. The effects of this canning have been studied during the measurement campaign by means of dedicated runs and were found to be negligible. The high-quality sample used for the measurement was instrumental for the accuracy of the results obtained.

In order to cover the same fraction of the neutron beam, all samples used in this experiment have been chosen with the same dimensions. A summary of the main features of the samples is reported in Table~\ref{tab:samples}.

The Data Acquisition System (DAQ) of the n\_TOF facility, presented in detail in Ref.~\cite{daq}, is fully digital, flexible and almost dead-time free. It has been designed based on 8-bit flash-ADCs with sampling rates of up to 2GHz and 8 Mbyte memory buffer. For the $^{238}$U(n,$\gamma$) measurement two channels of Acqiris-DC270 digitizers working at 500 MSamples/s are used per detector. This system corresponds to 96 ms long data buffers containing digitized signals of the C$_6$D$_6$ detectors for neutron energies between 0.02 eV and 20 GeV. 

\section{\label{sec:detperf}Operation and performances of the experimental apparatus}

To reach the aim of a precise and accurate $^{238}$U(n,$\gamma$) cross section the experimental setup has been precisely characterized, with particular attention to the stability and the performances of the scintillators. Moreover, in the present work we took advantage of the GEANT4~\cite{geant4} simulation toolkit for the study of the experimental area and of the detection system.

\subsection{Stability of the detectors}
The stability of the apparatus has been monitored with respect to the neutron flux and to the capture detectors. Concerning the neutron flux stability, we investigated the ratio between the counting rate of the silicon monitors and the number of protons derived from the current measurement via the pick-up signal. Runs were selected for further analysis only if this ratio was within 2\% of the mean. 

The stability of the two C$_6$D$_6$ detectors has been investigated by looking at the ratio of the C$_6$D$_6$ counting rate for different runs in the strongest resonances. This ratio has been recorded bunch per bunch, taking into account the corresponding number of protons. At most, the normalized number of entries per run deviates by about 6\% from the mean value, and those runs with percentage deviations greater than 3.5\% have been rejected in order to reduce the dispersion of the data set. With this constraint the distribution of the deviation was less than 2\% for 90\% of the runs. By these conditions we rejected in total 1.5\% of the data from the Bicron and 5\% from the FZK detector, not significantly affecting the statistics of the measurement. More details can be found in Ref.~\cite{Ming_procND}.

\subsection{Amplitude to deposited energy calibration}
As the PHWT requires the accurate energy calibration of the capture detectors, a careful study of the calibrations between the flash-ADC channels and the deposited energy has been performed on a weekly basis using three standard $\gamma$-ray sources: $^{137}$Cs (661.7 keV), $^{88}$Y (898 keV and 1.836 MeV) and Am/Be (4.44 MeV). The peaks in the amplitude spectra, corresponding to the respective Compton edge of the different $\gamma$ rays, are broadened by the detector resolution. This effect needs to be taken into account when extracting the channel numbers associated to the Compton edges, in order to properly calibrate the flash-ADC channels. To this purpose, the GEANT4 simulations of the detector response have been broadened to match the measured spectra. 

\begin{figure}[t]
\includegraphics[width=0.8\columnwidth]{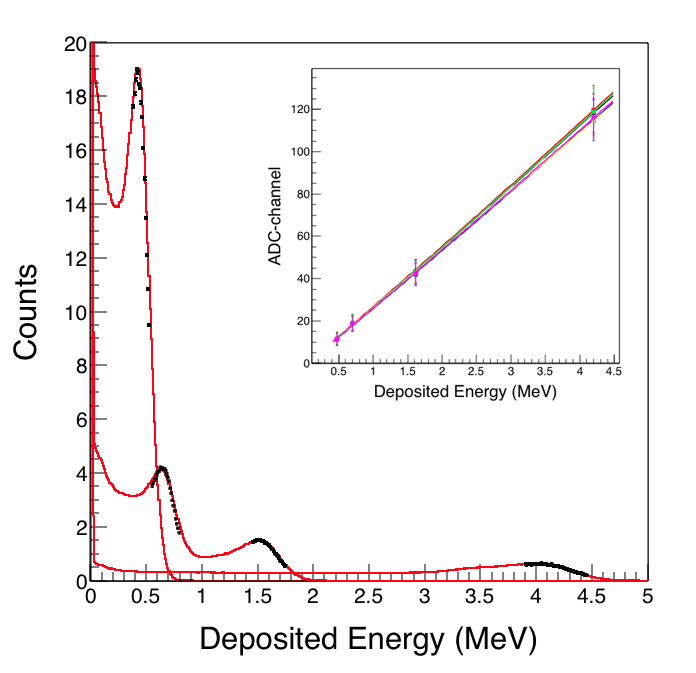}
\caption{\label{fig:calib}(Color online) Calibrated amplitude spectra for $^{137}$Cs ($E_\gamma=661.7$ keV), $^{88}$Y ($E_\gamma=898$ keV and $E_\gamma=1.836$ MeV) and Am/Be ($E_\gamma=4.44$ MeV) $\gamma$-ray sources. The red lines correspond to the broadened simulations, while the black dots are the experimental points. In the inset the different curves for the channel to deposited-energy calibration are shown.}
\end{figure}

This procedure is important to further check the stability of the detector gain over the whole campaign. Comparing the different calibration spectra for each source we actually noticed a small variation of the gain, which has been adjusted by applying different calibration curves. In Fig.~\ref{fig:calib} the Monte Carlo (MC) simulations of the detector response, broadened by the experimental resolution, are plotted together with experimental data. In the inset of Fig.~\ref{fig:calib} the different calibration lines are shown. 

For the $^{238}$U(n,$\gamma$) measurement the thresholds used for the analysis have been chosen to be $E_\text{dep}^\text{min} = 0.250$ MeV and $E_\text{dep}^\text{max} = 5.53$ MeV, corresponding to the Compton edge of $\gamma$-ray energies of 407 keV and 5.77 MeV, respectively. The upper threshold exceeds the neutron separation energy of $^{239}$U, $S_n = 4.806$ MeV, by 20\% to account for the resolution broadening.

\subsection{Weighting functions}
The PHWT technique requires a mathematical manipulation of the response function of the detection system to achieve proportionality between detection efficiency and deposited energy. To this purpose a weighting function WF($E_d$) is defined, and the detector response for different $\gamma$-ray energies, which depends on $\gamma$-ray transport both in the detection system and within the sample itself, has been simulated by means of the GEANT4 Monte Carlo simulation code. In these simulations, the $\gamma$ rays are emitted randomly within the sample following the Gaussian xy-distribution of the neutron beam profile, z being the direction of the neutron beam. 

The neutron transport within the sample can affect the gamma-ray spectrum recorded by the detector and becomes significant only for high values of the product $n\sigma_\text{tot}$, where $n$ is the areal density in atoms/barn and $\sigma_\text{tot}$ is the total cross section. In the present case, where the $^{238}$U sample has an areal density $n = 9.56\times 10^{-4}$ atoms/barn, this condition is fulfilled only near the peak-energy of the first three s-wave resonances. Since it is not possible to account for this effect, which depends on the cross section to be measured, we produced two different weighting functions for each detector, one considering an homogeneous distribution in the z-direction of the $\gamma$-ray emission within the sample, i.e. valid everywhere except for the saturated resonances, and another with an exponential attenuation, valid only in the saturation region of the resonances. The first WF has been used for the resonance shape analysis (RSA) and the analysis of the unresolved resonance region, as well as to weight the background counting rate. The second WF has been applied for the extraction of the normalization factor through the saturated resonance technique~\cite{Macklin_SR} (see Section~\ref{sec:Y_norm}).

The polynomial dependence of the weighting function WF on the energy deposited by $\gamma$ rays is determined by a least-squares fit to a number of $\gamma$-ray responses in the energy range of interest, i.e. from 250 keV up to 10 MeV. In the calculation the loss of counts due to the discrimination level has been taken into account as discussed in Ref.~\cite{Borella_C6D6}.

\section{\label{sec:datared}Data Reduction}
The experimental capture yield has been extracted dividing the weighted counts $C_w$, after correction for the dead time and subtraction of the corresponding background $B_w$, by the incident neutron fluence $\phi_n$~\cite{TOF_review}: 
\begin{equation}\label{eq:yield}
Y_{exp}(E_n) = \frac{N}{S_n + E_n \frac{A}{A+1}} \frac{C_w(E_n) - B_w(E_n)}{\phi_n(E_n)},
\end{equation}

where $N$ is the normalization factor, $E_n$ the energy of the incident neutron, $S_n$ the neutron separation energy of the compound nucleus and $A$ the mass number of the target nucleus. 

To achieve proportionality between deposited energy and detection efficiency by the PHWT, means also to take the efficiency of the two C$_6$D$_6$ scintillators for capture events into account. Therefore the normalization factor $N$ includes the effective area of the $^{238}$U sample intercepted by the neutron beam, which is $\sim 97\%$, and the absolute value of the incident neutron flux. 

\subsection{Time-of-flight to energy calibration}
The neutron velocity and therefore the kinetic energy were derived from the measured time-of-flight and the flight-path length, which is nominally 185 m. The effective flight path length $L(E)$ is not a fixed value but depends on the neutron energy according to the neutron flight time distribution in the spallation target and moderator assembly. This energy dependent distribution is also known as the resolution function (RF). For a given energy $E$, the effective flight path can be split up as the sum of a fixed flight path length $L_0$ and an additional component $\lambda(E)$, which corresponds to the expectation value of the equivalent distance for a neutron of energy $E$. The value of $L_0$, valid in combination with the used RF obtained by simulations, was fitted with the multilevel multichannel $R$-matrix code SAMMY~\cite{Larson_SAMMY} using the first 5 resonances and their corresponding energies. It is important to notice that in this low energy region $\lambda$ can be considered as a constant.

The procedure has been verified with the well known low-energy s-wave resonances of $^{197}$Au as in Ref.~\cite{Au_massimi}.

\subsection{Background evaluation and subtraction}
A precise background subtraction is fundamental for the accuracy desired for this work, as outlined in Ref.~\cite{Migrone_INPC}. Two different methods were considered for the study of the background, in order to assess its contribution and its uncertainty. Both methods are based on the identification of the individual background components, either determined through dedicated measurements or extrapolated from background measurements with the use of MC simulations. 

\begin{figure}[t]
\includegraphics[width=\columnwidth]{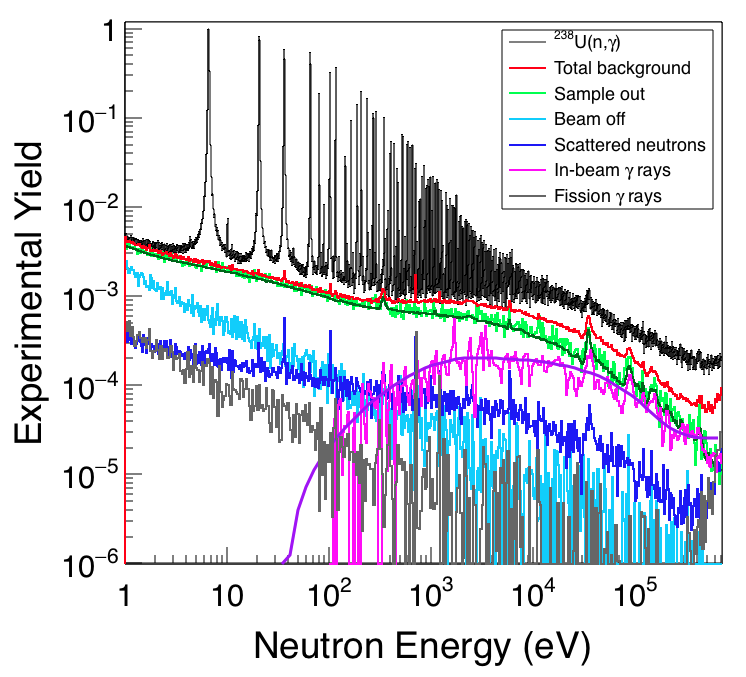}
\caption{\label{fig:Y_meas}(Color online) $^{238}$U(n,$\gamma$) capture yield compared with the total background and its individual components discussed in the text. The analytical functions which reproduce the beam-related background and the contribution of in-beam $\gamma$-rays are superimposed to the measured components.}
\end{figure}

In Fig.~\ref{fig:Y_meas} the $^{238}$U capture yield is shown together with the background decomposed into the different contributions. The total background has been calculated as
\begin{equation} \label{eq_bkg}
B = C_\text{BO}^U + \left(F_\text{EF} - C_\text{BO}^\text{EF} \right) + f_\text{inB}F_{\gamma \text{-inB}} + S_\text{NS} + S_\text{fiss}.
\end{equation}

The time independent source of background $C_\text{BO}$ comes from natural and sample radioactivity, and from air activation. Because this contribution turned out to be sensitive to the position of the sample on the vertical ladder of the sample changer, it has been routinely studied with different beam-off measurements corresponding to the different positions of the samples. This contribution dominates for low energies up to few eV.

The main component of the beam-related background has been evaluated without any sample in beam, in order to take all sources of background in the experimental area into account that are not related to the presence of the sample. In order to avoid statistical fluctuations, the resulting shape, as a function of TOF, has been fitted with an analytical function based on a sum of parametrized exponentials ($F_\text{EF}$ in Eq.~\ref{eq_bkg}), and an uncertainty of 5\% has been associated to that function. 

The beam-related background due to in-beam $\gamma$ rays, although strongly suppressed by the borated water moderator, plays an important role in the keV energy region. Its shape can be deduced by means of a n+$^{nat}$Pb measurement properly subtracted by beam-off and sample-out contributions. The difference in areal density between uranium and lead sample as well as the different charge number were taken into account as a scaling factor ($f_\text{inB}$ in Eq.~\ref{eq_bkg}), obtained through GEANT4 Monte Carlo simulations using the implemented geometry for the calculation of the weighting functions. The initial in-beam $\gamma$-ray distribution has been provided by Monte Carlo FLUKA simulations (see Ref.~\cite{Gue_report} for more details about the simulated $\gamma$-ray spectra in the experimental area). To avoid statistical fluctuations the analytical form of this background has been calculated and was fitted to the measured spectrum of Pb after applying the scaling factor obtained with the simulations ($F_{\gamma \text{-inB}}$ in Eq.~\ref{eq_bkg}).

Another source of background for (n,$\gamma$) measurements is related to the detection of $\gamma$-rays coming from sample-scattered neutrons captured in the ambient material. This background follows the same energy dependence as the true capture events and may therefore compromise the analysis of resonances. To properly evaluate it, GEANT4 simulations have been performed with a complete description of the geometry of the experimental area. All the details of the procedure can be found in Ref.~\cite{Zugec_NS}. The output of the simulation has been analyzed using the same conditions as for the experimental data ($S_\text{NS}$ in Eq.~\ref{eq_bkg}).

The same simulations have been used to evaluate the contribution of $\gamma$ rays coming from fission events ($S_\text{fiss}$ in Eq.~\ref{eq_bkg}). They can be due to prompt $\gamma$-rays and $\gamma$-rays from the decay of fission-fragment. We used the models and data sets as reported in Ref.~\cite{Zugec_NS}, which for $^{238}$U provide data only up to 20 MeV. Therefore we had to account for the missing part of the neutron spectrum with a scaling procedure, assuming that the shape of the $\gamma$ background coming from fission events does not change with increasing energy. We multiplied the reference $^{238}$U(n,$f$) cross section from Ref. \cite{IAEA_ref} with the simulated n\_TOF neutron flux to obtain an estimated count rate, and we scaled the simulated background for the ratio between the integral of the expected count rate up to 20 MeV and up to 1 GeV. 

The overall background is dominated by the general background, independent of the sample itself. Only in the unresolved resonance region (URR) above about 10 keV the in-beam $\gamma$-rays start to contribute a comparable fraction. All other background components are of minor importance. The favorable signal/background ratio in the resolved resonance region (RRR) is shrinking at higher energies but remains always better than 2:1 throughout the URR.

\begin{figure}[b]
\includegraphics[width=\columnwidth]{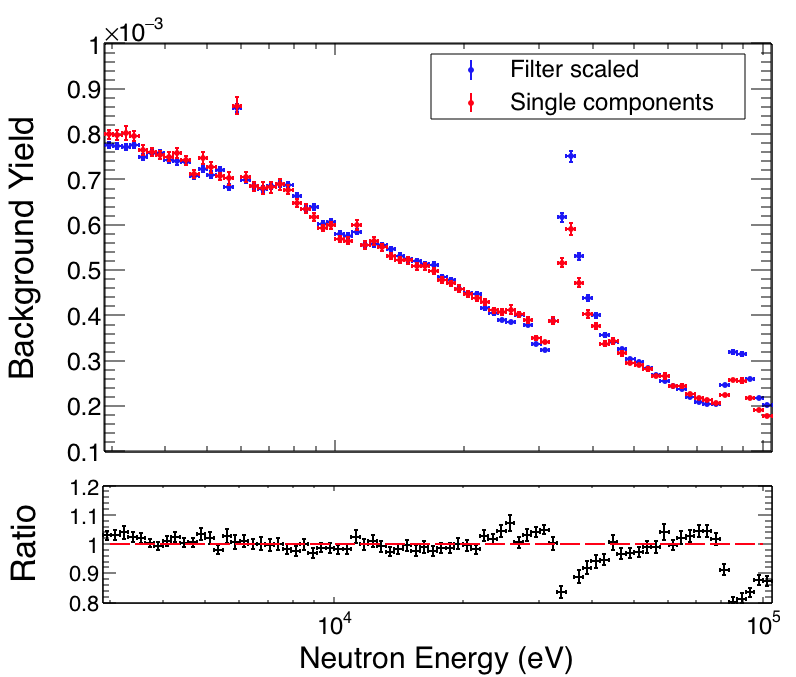}
\caption{\label{fig:diff_bkg}(Color online) Background yield in the URR obtained by summing the individual background components shown in Fig.~\ref{fig:Y_meas} (red) and by using the black-resonance filters in the neutron beam (blue). The bottom panel shows that the ratio deviates by less than 5\% from unity, except in the neighbourhood of the two big Al resonances at 34.23 and 84.27 keV, because the amount of aluminum in the windows of the beam line was not precisely known.}
\end{figure}

To validate the level of the background in the URR we exploited measurements with Ag, W, Co and Al black resonance filters in beam~\cite{TOF_review}. The shape of the time-dependent background has been evaluated in a n+$^{nat}$Pb measurement, and the beam-off contribution was added in order to obtain the complete background function. The resulting background has been properly scaled to reproduce the background during the measurement according to the procedure introduced by Aerts \textit{et al.}~\cite{Aerts_bkgURR}. In particular, the total background, in counts per log-equidistant time-of-flight bin, can be reproduced by the function:
\begin{equation}\label{eq:Ubkg}
B(t) = a t^b + \frac{k_1}{k_2(t)} \; Pb(t),
\end{equation}
where $a t^b$ reproduces the constant beam-off background, with $b$ fixed to 1 because of the units of $B(t)$. To take into account the attenuation of both the neutron beam and the in-beam photons, two scaling factors have been identified: a time-independent factor $k_1$ to scale the contribution of n+$^{nat}$Pb without filters to the n+$^{238}$U with filters, and a time-dependent factor $k_2(t)$ to take the attenuation of the in-beam $\gamma$-rays by the filters into account. The time-dependence reflects the fact that the relative intensities of the in-beam $\gamma$ rays change with time-of-flight.

As shown in Fig.~\ref{fig:diff_bkg}, the background yields agree agree for $3<E_n<100$ keV within less than 5\% except for energies in the neighbourhood of the two strong Al resonances at 34.23 and 84.27 keV. For neutron energies above 100 keV, it is impossible to determine the background with the Al filter reliably because of the increasing resonance structure of the Al cross section.

\subsection{Capture yield normalization}\label{sec:Y_norm}
The experimental capture yield has been internally normalized to the peaks of the first three well-isolated saturated resonances at 6.67 eV, 20.9 eV and 36.7 eV. Since $n \sigma_{tot} \gg 1$ at the resonance peak, all the incident neutrons with energies close to the resonance interact with the sample and therefore the effects of neutron transport cannot be neglected in the region of a saturated resonance. For this reason the weighting function that includes an exponential attenuation within the sample has been used. 

A spectrum with a weighting function was also calculated for homogeneous $\gamma$-ray production inside the sample, corresponding to $n\sigma_{tot} \ll 1$. The ratio of the weighted spectra obtained with the two weighting functions deviated from unity by less than 0.5\%. For this reason we neglected the energy-dependent yield correction as in Ref. \cite{Norm_Th}.
   
The normalization factor $N$ in Eq.~\ref{eq:yield} has been extracted via the saturated resonance method, evaluating the expected capture yield through a least-squares adjustment of the experimental data with the SAMMY code~\cite{Larson_SAMMY}. Because an internal normalization is applied, i.e. exploiting a resonance of the nucleus under investigation, the systematic uncertainties due to changes in the experimental conditions are significantly reduced.

The two detectors have been treated separately, and for each one the final normalization factors have been chosen as the average of the three resonances. The mean deviation of these factors of 1\% has been adopted as the normalization uncertainty.

\subsection{Discussion of uncertainties}
\begin{table*}
\caption{\label{tab:sum_unc}Summary of the correlated uncertainties in the $^{238}$U(n,$\gamma$) cross section measurement. The uncertainties on the neutron flux shape are from Ref~\cite{flux}.}
\begin{ruledtabular}
\begin{tabular}{llll}
 \multicolumn{1}{l}{Source of uncertainty} &  \multicolumn{1}{l}{Uncertainty RRR}   & \multicolumn{1}{l}{Uncertainty URR} & \multicolumn{1}{l}{Uncertainty URR }   \\
&   & \multicolumn{1}{l}{($3<E_n<100$ keV)} & \multicolumn{1}{l}{($E_n>100$ keV)}   \\
      \hline
Sample mass & 0.1\%   & 0.1\% & 0.1\%  \\
Neutron flux - shape & 1-2\% & 4-5\% & $\sim2\%$  \\
Normalization & $1\%$   & $1\%$ & $1\%$  \\
Background subtraction & 1-2\% & 3\% & $\sim7\%$  \\
\hline
Total & 2-3\% & 5-6\%  & 8\%  \\
\end{tabular}
\end{ruledtabular}
\end{table*}
The total uncertainty in the $^{238}$U(n,$\gamma$) cross section is a combination of several components related to the sample characteristics and the analysis procedure. Along with the uncorrelated uncertainty due to the counting statistics, correlated uncertainties are involved as well.

The uncertainty related to sample characterization is almost negligible since  the sample mass has been determined at the EC-JRC with an accuracy of about 0.1\%. 

An important source of correlated uncertainty is the yield normalization. As discussed in Subsection~\ref{sec:Y_norm}, the uncertainty in the normalization factor is reduced to less than 1\% by the internal normalization. This correlated uncertainty component has to be combined with the uncertainty of the neutron flux shape between 1 and 5\%, depending on neutron energy~\cite{flux}.

The uncertainty related to the background subtraction propagates to the final capture yield in correlation to the signal-to-background ratio, and it depends on the energy range considered. Within the RRR the signal to background ratio varies from a factor of 4 to a factor of 100, depending on the strength of the resonance. In the URR the signal to background ratio is about a factor of 2, resulting in a ~3\% uncertainty of the background subtraction for 3 $< E_n < 100$ keV. 

For higher energies from 100 to 700 keV the level of the background strongly depends on the $\gamma$-ray component, which has been estimated from MC simulations. An uncertainty of $\sim 7\%$ has been attributed to the background level for this energy range. The estimation of the uncertainty has been based on the comparison between the GEANT4 full simulations (i.e. considering all background contributions) and the experimental data for $^{197}$Au. The agreement between the two, within a few percent, provided confirmation of the level of accuracy of the simulated background.

In Table~\ref{tab:sum_unc} all the correlated uncertainties are listed for the different energy ranges. 

\section{Results}

\subsection{Resolved Resonance Region}
The energy resolution of the n\_TOF spectrometer gives an upper energy limit of 20 keV to the RRR, since the experimental broadening becomes comparable with the level spacing. However, due to the decrease of the signal-to-background ratio with neutron energy together with the limited counting statistics, resonances could be reliably resolved only up to 3 keV. 

\begin{figure}[b]
\includegraphics[width=\columnwidth]{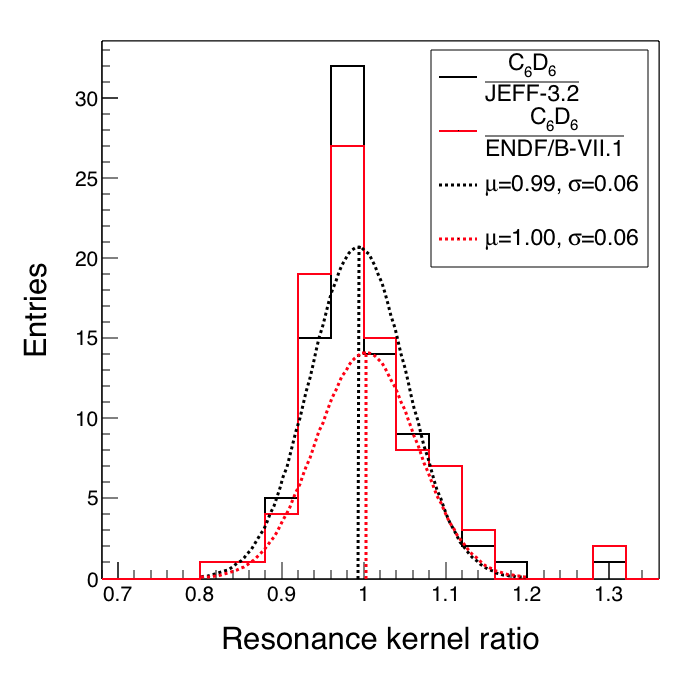}
\caption{\label{fig:stat_k}(Color online) Statistical distribution of resonance kernel ratios of this work over JEFF-3.2 and this work over ENDF/B-VII.1. The gaussian best-fit curve is also plotted as a dashed line.}
\end{figure}

\begin{figure*}
\includegraphics[width=1.5\columnwidth]{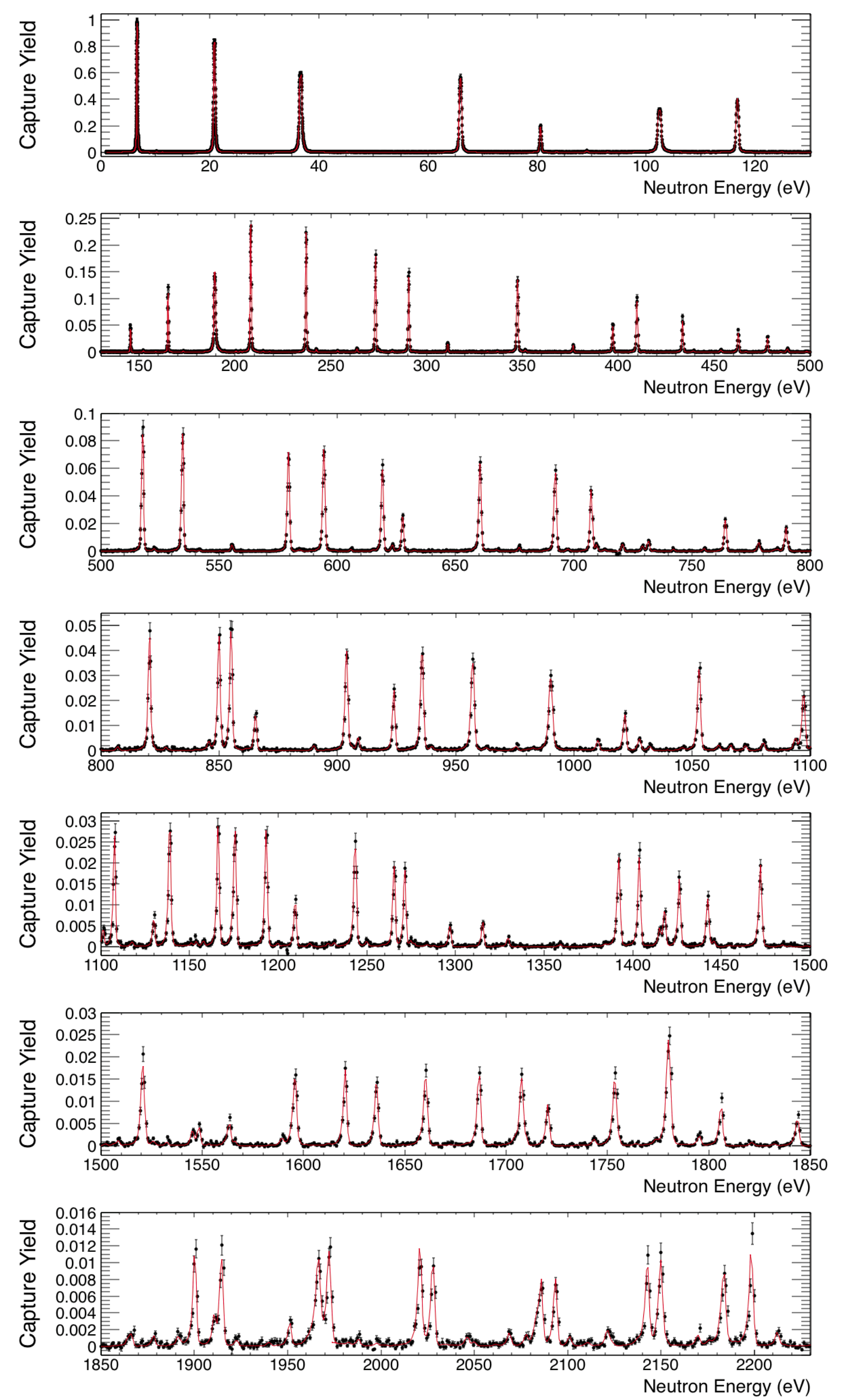}
\caption{\label{fig:sammy_fit}(Color online) Experimental capture yield for $^{238}$U(n,$\gamma$) with a resolution of 5000 bin/decade. The SAMMY  calculation based on JEFF-3.2~\cite{Jeff-3.2} resonance parameters is represented by the solid line.}
\end{figure*}

Within this energy region a resonance shape analysis has been performed by means of the SAMMY code~\cite{Larson_SAMMY}. The theoretical reaction yield is obtained through the Reich-Moore approximation of the multi-level R-matrix formalism, and convoluted with the experimental resolution. This latter is implemented in SAMMY via a numerical description derived from Monte Carlo FLUKA simulations. Moreover, the code includes the corrections for multiple-scattering effects and uses the free-gas model to take the Doppler broadening due to thermal motion of atoms inside the sample for an effective temperature of 305 K into account, according to the room temperature which was continuously monitored during the measurement \cite{TOF_review}. 

The resonance parameters (i.e. the capture width $\Gamma_\gamma$ and the neutron width $\Gamma_n$) are determined by a least-square fit to the experimental data, while the resonance spins are taken from literature. In this work, starting from initial values taken from the JEFF-3.2 library~\cite{Jeff-3.2}, only the smaller parameter, either $\Gamma_\gamma$ or $\Gamma_n$, has been left free in the fitting procedure. In a few cases both partial widths were left free to improve the description of the resonance shape analysis. 

\begin{figure*}
\centering
\includegraphics[width=1.25\columnwidth]{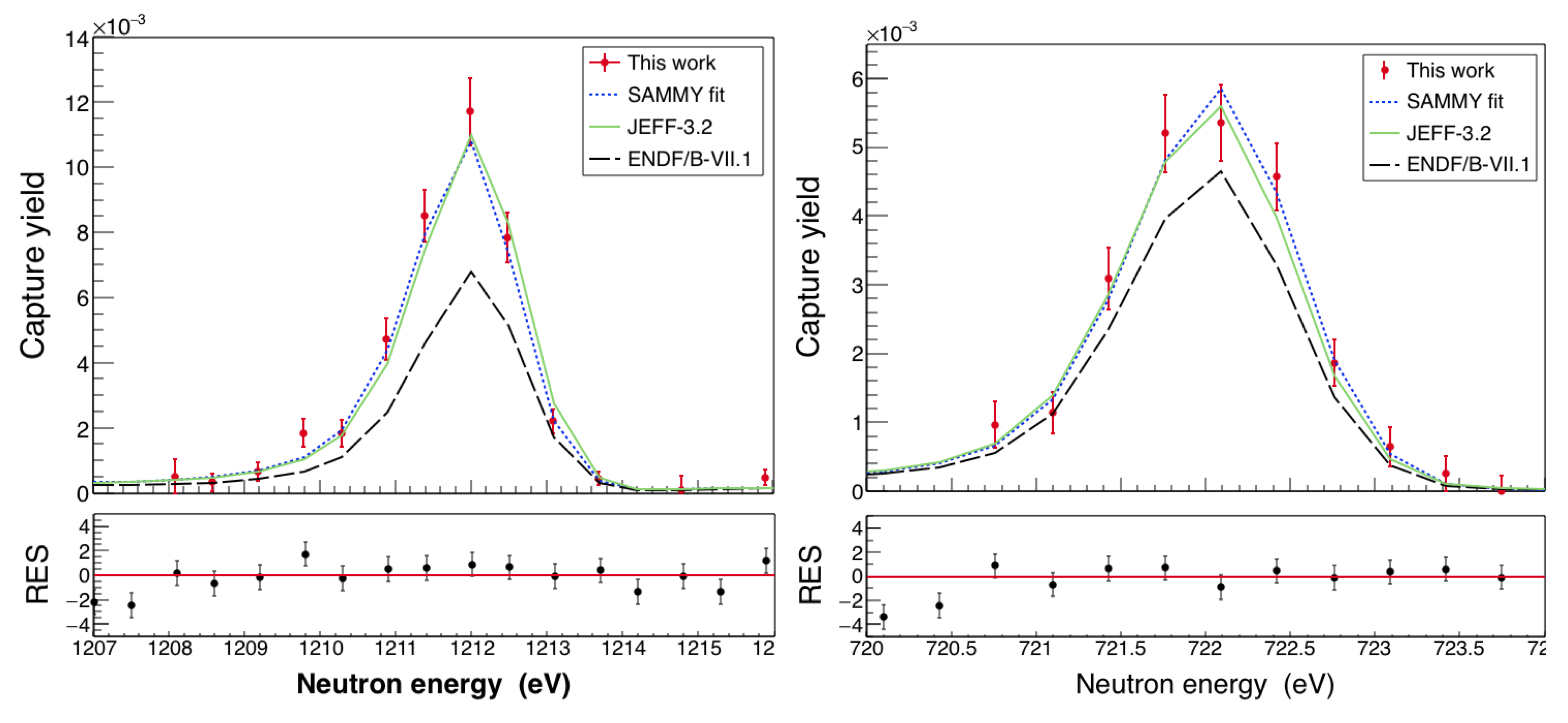}
\caption{\label{fig:res_721}(Color online) Resonances at $E_R = 721.68$ eV and $E_R = 1211.31$ eV. Data from this work are red points, and the SAMMY best fit is shown as a dotted blue line. Calculations performed using JEFF-3.2 and ENDF/B-VII.1 resonance parameters are shown as continuous-green and dotted-black lines respectively. The bottom panel shows the residuals of the fit.}
\end{figure*}

\begin{figure*}
   \centering
    \includegraphics[width=1.3\columnwidth]{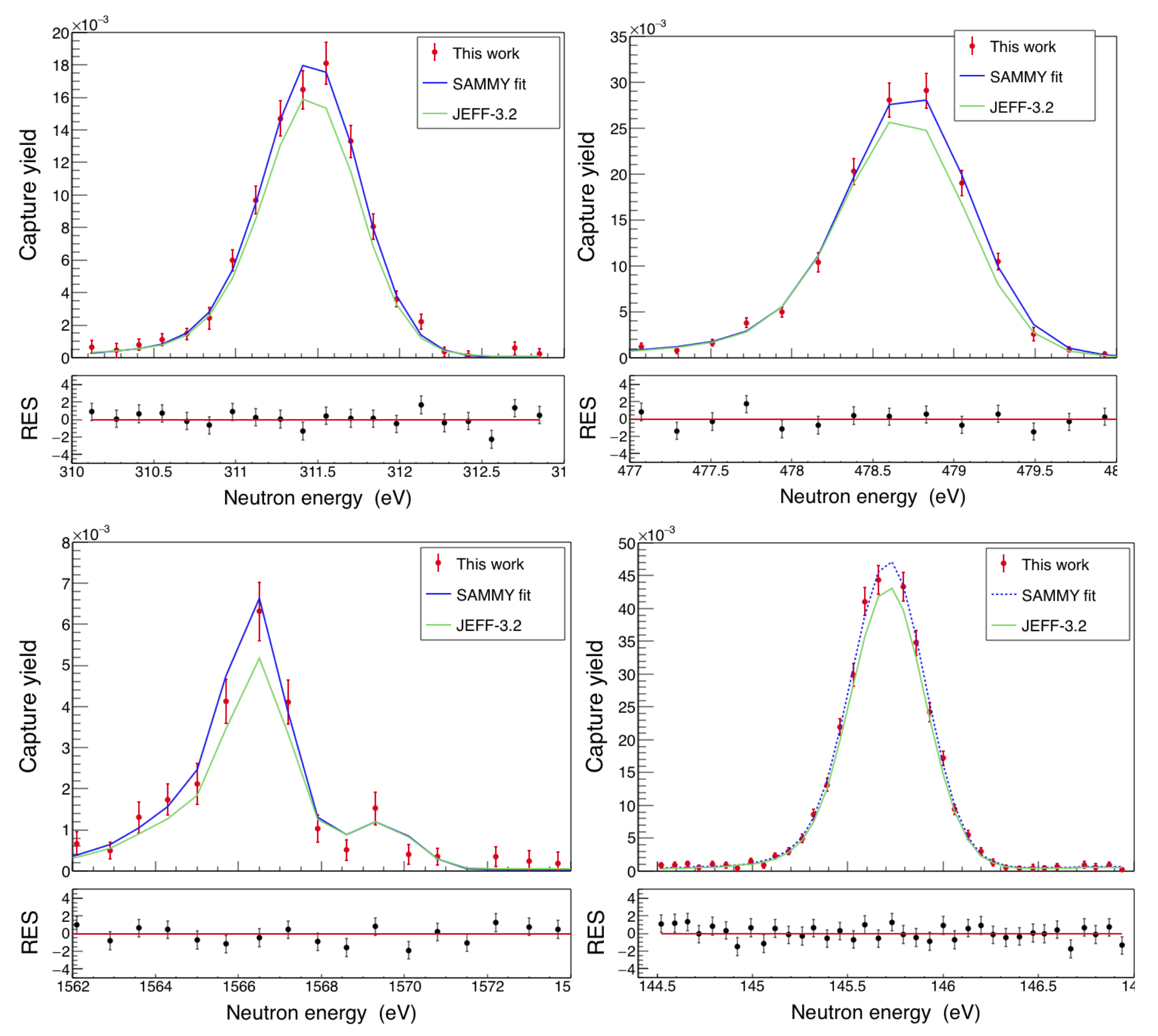}
 \caption{\label{fig:res_fit}(Color online) Resonances at $E_R = 145.66$ eV (top left),  $E_R = 311.35$ eV (top right),  $E_R =488.89$ eV (bottom left) and $E_R = 1565.54$ eV (bottom right). Data from this work are red points, and the SAMMY best fit is shown as a dotted blue line. Calculation performed using JEFF-3.2 resonance parameters is shown as a green line for comparison. The bottom panel shows the residuals of the fit.}
\end{figure*}

The results from this work have been compared with the evaluated data libraries JEFF-3.2~\cite{Jeff-3.2} and ENDF/B-VII.1~\cite{ENDF-7.1}. The ratio of kernels (defined as $\kappa = g\Gamma_n\Gamma_\gamma/(\Gamma_n+\Gamma_\gamma)$) from this work to the ones obtained from evaluated parameters from JEFF-3.2 and ENDF/B-VII.1 has been studied as a function of resonance energy to test the accuracy of the measurement. Because of the large number of measurements carried out so far and the effort spent in producing an accurate $^{238}$U(n,$\gamma$) evaluation, significant deviation from the evaluated data are not expected. As can be seen in Fig.~\ref{fig:stat_k}, the statistical distributions are found to be Gaussians centered at 0.99 and 1.00 respectively, both with a standard deviation of 0.06, showing the excellent agreement between data from this work and evaluated libraries. The measured capture yield together with the SAMMY best-fit curve is shown in Fig.~\ref{fig:sammy_fit}. 

Below 1.5 keV, the JEFF-3.2 and ENDF-B/VII.1 libraries present sizable differences for some resonances. In this energy region, there are two resonances from this work, at 721.68 eV and 1211.31 eV, respectively, which strongly differ from those indicated in the ENDF/B-VII.1 library, while they are in good agreement with JEFF-3.2. In particular, the ENDF/B-VII.1 evaluation gives $\Gamma_\gamma = 3.15$ meV and $6.6$ meV for these resonances, whereas JEFF-3.2 reports $\Gamma_\gamma = 23.0$ meV and $\Gamma_\gamma = 17.55$ meV, respectively. As can be seen in Fig.~\ref{fig:res_721}, the partial widths obtained from this work are in agreement with the ones reported by JEFF-3.2. Nevertheless, the difference with ENDF/VII.1 can be explained to some extent with a different treatment of the fission contribution \cite{Difilippo, Auchampaugh}.

Residual systematic effects related to neutron scattering or to the resonance strength have been evaluated studying kernel ratios as a function of the ratio $g\Gamma_n/\Gamma_\gamma$ and of the resonance kernel itself. In the former case, no evidence of systematic effects are visible as expected from the optimized detection set-up used. Regarding effects due to the resonance strength, it has been seen that for weak resonances, characterized by very small kernels ($\kappa \leq 2.3$ meV), resonance parameters obtained from this work are systematically larger than the evaluated ones. Despite the low strength of these resonances, three of them are in the low energy region, $E_R = 145.66,~ 311.35, ~488.89$ eV, where the statistics is very high. The fitting procedure is therefore quite accurate, and Fig.~\ref{fig:res_fit} shows that data from this work are actually higher than what is predicted by using the JEFF-3.2. parameters The same conclusion can be drawn for the resonance at $E_R = 1565.54$ eV, a stronger resonance, which is $\sim 27\%$ higher than predicted from JEFF-3.2 parameters. The kernels of critical resonances obtained from this work are compared with the evaluated parameters in Table~\ref{tab:kernels}.

\begin{table}[!h]
\caption{\label{tab:kernels}Resonance kernels ($\kappa$) for $^{238}$U+n calculated from parameters obtained from this work, and from ENDF/B-VII.1 and JEFF-3.2 libraries.}
\begin{ruledtabular}
\begin{tabular}{ccccc}
 & \multicolumn{2}{c}{This work} & ENDF/B-VII.1 & JEFF-3.2 \\
 $E_R$ (eV) & $\kappa$ (meV) & $\Delta \kappa/ \kappa$ (\%)& $\kappa$ (meV)& $\kappa$ (meV) \\
 \hline
145.66	 & $0.93  \pm 0.02$	 & 2	 	 & 0.85	 & 0.85 \\
311.35	 & $1.14  \pm 0.04$	 & 3	 	 & 1.01	 & 1.00 \\
488.89	 & $1.0 \pm 0.4$	 & 44	 	 & 0.83	 & 0.83 \\
721.68	 & $1.45 \pm 0.09$	 & 6	 	 & 1.12	 & 1.36 \\
1211.32	 & $6.5 \pm 4$		 & 61	 	 & 4.01	 & 6.54 \\
1565.54	 & $6.1 \pm 0.7$	 & 12	 	 & 4.71	 & 4.71 \\
\end{tabular}
\end{ruledtabular}
\end{table}

\subsection{Unresolved resonance region}
As the experimental conditions are limiting the energy region where the cross section can be analyzed through a RSA to about 3 keV, the capture cross-section at energies up to 700 keV is obtained from the measured yield by applying a correction factor determined from average parameters.

To extract the cross section the thin-sample approximation $Y \approx n\sigma_\gamma$  is used. At high energies, in fact, the total cross section becomes quite low ($\sigma_\text{tot} \sim 1$ barn) sto that the areal density $n=9.56\times 10^{-4}$ at/barn is sufficiently low to fulfil the condition $n \sigma_\text{tot} \ll 1$. 

\begin{figure}
   \centering
    \includegraphics[width=\columnwidth]{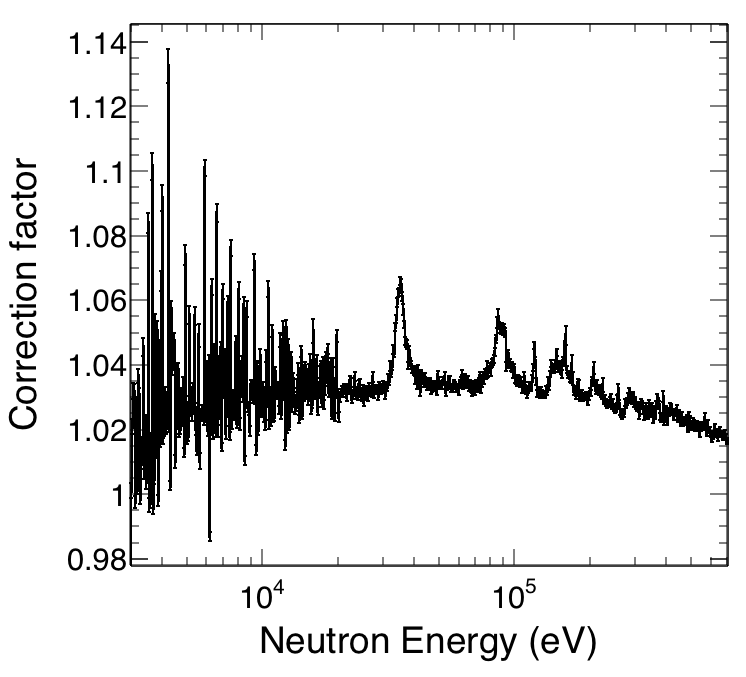}
 \caption{\label{fig:corr_factor} Correction factor for the sample-related effects, i.e. self-shielding and multiple scattering followed by capture, obtained from MCNP6 simulations.}
\end{figure}

The sample-related effects, i.e. self-shielding and multiple scattering followed by capture, are taken into account applying a correction factor obtained from Monte Carlo simulations. This correction factor $C_f$ relates the energy-averaged capture cross section $\overline{\sigma}_\gamma$ to the measured energy-averaged capture yield $\overline{Y}_c$ by:
\begin{equation} 
\overline{\sigma}_\gamma = \frac{\overline{Y}_c}{n \times C_f}.
\end{equation}
The full geometry of the sample has been modelled, including the protective aluminum and Kapton cover foils of total thickness 60 $\mu$m and 75 $\mu$m respectively. We used MCNP6 \cite{MCNP} to generate neutrons with a Gaussian spatial beam profile impinging on the $^{238}$U sample and to tally the time when a capture reaction occurred in the $^{238}$U material. This time can then be converted to an equivalent neutron energy. From the number of capture reactions and the number of incident neutrons the quantity $\overline{Y}_c$ can be calculated. This observable as a function of time is the most adequate approach to obtain a quantity comparable to a time-of-flight measurement, and is also used for multiple scattering corrections in the resolved resonance region \cite{Dagan_MS}. The evaluated nuclear data library ENDF/B-VII.1 was used for all nuclei in the simulations. By trying different variations we found that the difference between a uniform or Gaussian beam profile is negligible. On the other hand, the presence of aluminum has a non-negligible effect as can be seen from the peaks corresponding to Al resonances in Fig.~\ref{fig:corr_factor}. The results were confirmed by simulations with the code GEANT4. The factor $C_f$ shown in Fig.~\ref{fig:corr_factor} was applied to the background-corrected experimental capture yield to obtain the average capture cross section $\overline{\sigma}_\gamma$.

Table~\ref{tab:URR} summarizes the average capture cross-section together with the total uncertainties derived from the data analyzed in this work. Correlated uncertainties are separated for the different components listed in Table~\ref{tab:sum_unc}.

\begin{table*}
\caption{\label{tab:URR} Average capture cross-section ($\overline{\sigma}_\gamma$) and total uncertainty from this work, for neutron energy bins between $E_l$ and $E_h$. The statistical uncertainty is listed in column 4, and the correlated uncertainties of Table~\ref{tab:sum_unc} are listed in columns 6 to 9. The correction factor of Fig.~\ref{fig:corr_factor} is given in column 5.}
\begin{ruledtabular}
\begin{tabular}{llllllllll}
$E_{l} $(keV) & $E_{h}$ (keV) & $\overline{\sigma}_\gamma$ (mb) & $u_{\overline{\sigma}_\gamma}$ (mb) & $CF$ & $u_\text{mass}$ (mb) & $u_\text{flux}$ (mb) & $u_\text{norm}$ (mb) & $u_\text{bkg}$ (mb) \\
\hline
3  &  6  &  1026  &  8.9636  &  1.0309  &  1.0258  &  46.1588  &  10.2575  &  30.7725\\ 
6  &  9  &  811  &  10.3406  &  1.0350  &  0.8107  &  36.4824  &  8.1072  &  24.3216\\ 
9  &  12  &  651  &  10.3441  &  1.0330  &  0.6507  &  29.2825  &  6.5072  &  19.5217\\ 
12  &  15  &  634  &  11.4595  &  1.0330  &  0.6342  &  28.5394  &  6.3421  &  19.0263\\ 
15  &  18  &  585  &  11.8854  &  1.0346  &  0.5846  &  26.3088  &  5.8464  &  17.5392\\ 
18  &  21  &  541  &  12.0580  &  1.0333  &  0.5406  &  24.3250  &  5.4055  &  16.2166\\ 
21  &  24  &  461  &  11.6140  &  1.0321  &  0.4606  &  20.7271  &  4.6060  &  13.8181\\ 
24  &  27  &  446  &  12.5137  &  1.0316  &  0.4462  &  20.0800  &  4.4622  &  13.3867\\ 
27  &  30  &  447  &  12.2569  &  1.0325  &  0.4470  &  20.1150  &  4.4700  &  13.4100\\ 
30  &  35  &  440  &  12.9528  &  1.0433  &  0.4403  &  19.8157  &  4.4035  &  13.2105\\ 
35  &  40  &  390  &  15.3661  &  1.0484  &  0.3898  &  17.5406  &  3.8979  &  11.6938\\ 
40  &  45  &  386  &  11.7462  &  1.0356  &  0.3860  &  17.3715  &  3.8603  &  11.5810\\ 
45  &  50  &  336  &  10.6194  &  1.0338  &  0.3355  &  15.0986  &  3.3553  &  10.0658\\ 
50  &  55  &  301  &  10.2271  &  1.0342  &  0.3008  &  13.5349  &  3.0077  &  9.0232\\ 
55  &  60  &  263  &  10.1233  &  1.0332  &  0.2626  &  11.8169  &  2.6260  &  7.8779\\ 
60  &  65  &  259  &  9.5294  &  1.0347  &  0.2593  &  11.6698  &  2.5933  &  7.7799\\ 
65  &  70  &  235  &  9.1976  &  1.0333  &  0.2353  &  10.5895  &  2.3532  &  7.0597\\ 
70  &  75  &  209  &  8.7905  &  1.0342  &  0.2086  &  9.3849  &  2.0855  &  6.2566\\ 
75  &  80  &  201  &  9.2020  &  1.0375  &  0.2014  &  9.0623  &  2.0138  &  6.0415\\ 
80  &  85  &  229  &  11.8405  &  1.0424  &  0.2290  &  10.3028  &  2.2895  &  6.8686\\ 
85  &  90  &  215  &  12.9878  &  1.0521  &  0.2149  &  9.6691  &  2.1487  &  6.4461\\ 
90  &  95  &  209  &  11.8490  &  1.0457  &  0.2092  &  9.4142  &  2.0920  &  6.2761\\ 
95  &  100  &  209  &  10.6563  &  1.0393  &  0.2088  &  9.3967  &  2.0881  &  6.2644\\ 
100  &  130  &  206  &  4.8759  &  1.0343  &  0.2064  &  4.1284  &  2.0642  &  14.4493\\ 
130  &  160  &  195  &  5.2822  &  1.0383  &  0.1951  &  3.9023  &  1.9511  &  13.6580\\ 
160  &  190  &  169  &  4.5366  &  1.0335  &  0.1689  &  3.3776  &  1.6888  &  11.8216\\ 
190  &  220  &  156  &  4.3657  &  1.0325  &  0.1564  &  3.1280  &  1.5640  &  10.9481\\ 
220  &  250  &  156  &  4.0589  &  1.0289  &  0.1557  &  3.1150  &  1.5575  &  10.9025\\ 
250  &  280  &  158  &  3.9849  &  1.0270  &  0.1578  &  3.1560  &  1.5780  &  11.0460\\ 
280  &  310  &  153  &  4.2810  &  1.0291  &  0.1532  &  3.0644  &  1.5322  &  10.7255\\ 
310  &  340  &  137  &  3.9519  &  1.0262  &  0.1368  &  2.7364  &  1.3682  &  9.5774\\ 
340  &  370  &  139  &  3.9561  &  1.0262  &  0.1386  &  2.7715  &  1.3858  &  9.7004\\ 
370  &  400  &  127  &  3.7996  &  1.0256  &  0.1271  &  2.5426  &  1.2713  &  8.8993\\ 
400  &  430  &  146  &  4.9321  &  1.0253  &  0.1462  &  2.9240  &  1.4620  &  10.2341\\ 
430  &  460  &  123  &  4.7144  &  1.0244  &  0.1234  &  2.4671  &  1.2335  &  8.6348\\ 
460  &  490  &  113  &  3.6950  &  1.0225  &  0.1127  &  2.2545  &  1.1273  &  7.8908\\ 
490  &  520  &  130  &  3.8603  &  1.0220  &  0.1303  &  2.6066  &  1.3033  &  9.1232\\ 
520  &  550  &  117  &  3.7326  &  1.0207  &  0.1170  &  2.3406  &  1.1703  &  8.1920\\ 
550  &  580  &  129  &  3.9317  &  1.0214  &  0.1288  &  2.5756  &  1.2878  &  9.0147\\ 
580  &  610  &  124  &  4.1569  &  1.0194  &  0.1243  &  2.4865  &  1.2432  &  8.7027\\ 
610  &  640  &  125  &  4.9586  &  1.0190  &  0.1245  &  2.4904  &  1.2452  &  8.7165\\ 
640  &  670  &  116  &  5.3016  &  1.0187  &  0.1164  &  2.3280  &  1.1640  &  8.1480\\ 
670  &  700  &  117  &  5.3735  &  1.0177  &  0.1170  &  2.3407  &  1.1703  &  8.1923\\ 
\end{tabular}
\end{ruledtabular}
\end{table*}

\begin{figure}
   \centering
    \includegraphics[width=\columnwidth]{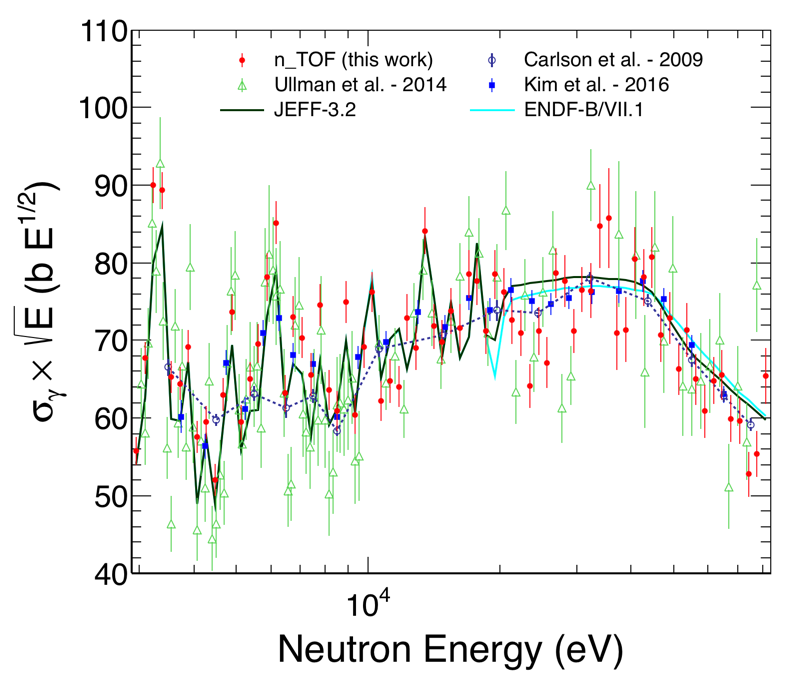}
 \caption{\label{fig:URR_1st}(Color online) $^{238}$U(n,$\gamma$) cross section from this work compared to the recently published data by Ullmann \textit{et al.}~\cite{Ullmann} and Kim \textit{et al.}~\cite{Kim}, and to the cross section recommended by Carlson \textit{et al.}~\cite{Carlson}. The evaluated cross sections from JEFF-3.2 and ENDF/B-VII.1 are plotted for comparison.}
\end{figure}

The $^{238}$U capture cross-section from this work has been compared to evaluated libraries and previous measurements (retrieved from the EXFOR database~\cite{exfor}). From 3 to 80 keV data from this work are plotted in Fig.~\ref{fig:URR_1st} with a resolution of 50 bin/decade (i.e. $dE/E = 2\%$). As can be seen, up to 20 keV they substantially agree with the recently published data by Ullmann \textit{et al.}~\cite{Ullmann} and Kim \textit{et al.}~\cite{Kim}, and with the cross section recommended by Carlson \textit{et al.}~\cite{Carlson} and by both JEFF-3.2 and ENDF-B/VII.1 evaluated libraries. For neutron energies from 20 to 80 keV data from this work stay slightly below the evaluated libraries and the other measurements.

\begin{figure}
   \centering
    \includegraphics[width=\columnwidth]{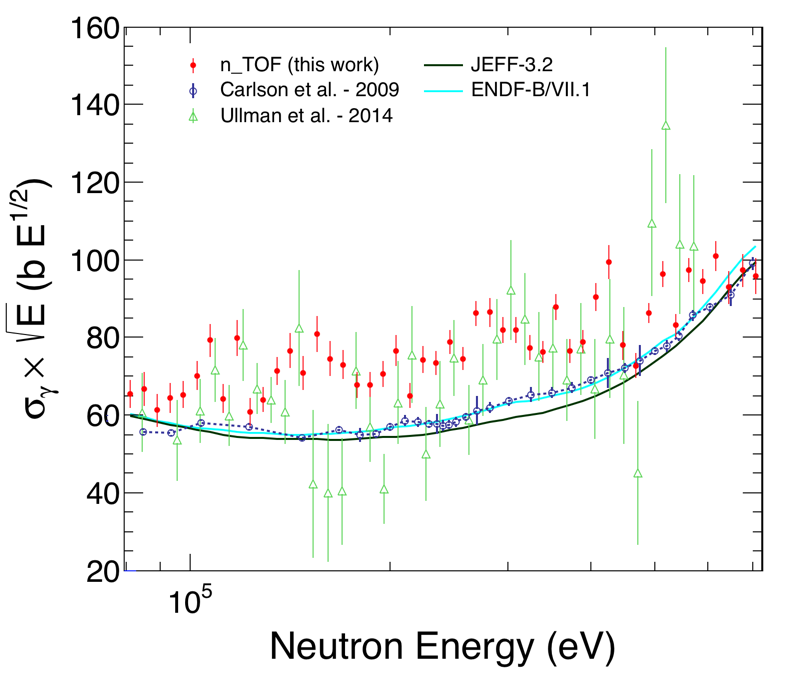}
 \caption{\label{fig:URR_2nd}(Color online) Measured $^{238}$U(n,$\gamma$) cross section in the energy range $80<E_n <700$ keV: data from this work compared to Ullmann \textit{et al.}~\cite{Ullmann} and to the cross section recommended by Carlson \textit{et al.}~\cite{Carlson}. Also shown are the evaluated cross sections from JEFF-3.2 and ENDF/B-VII.1.}
\end{figure}

For higher energies from 80 up to 700 keV, as shown in Fig.~\ref{fig:URR_2nd}, this work yields a cross section that is 15 to 25\% higher than the ENDF/B-VII.1 evaluation and the cross section of Carlson \textit{et al.}, and 21 to 32\% higher than the JEFF-3.2 evaluation. The measurement from Ullmann \textit{et al.} shows a fair agreement with the present work up to 630 keV.

Data from the present work and from Ullmann \textit{et al.} have been compared with the spectrum-averaged cross-section data from Wallner \textit{et al.} \cite{Wallner}, recently obtained exploiting a novel method for neutron-energy distributions with a mean of 25.3 keV and 426 keV.

For the approximate Maxwell-Boltzmann distribution at $kT=25.3$ keV the present analysis and the data by Ullmann \textit{et al.} differ from the results by Wallner \textit{et al.} only by 3 and 2\%, respectively, in agreement within the uncertainties. On the contrary, for the energy distribution peaked at 426 keV the present analysis and data by Ullmann \textit{et al.} yield results that are 16 and $12\%$ higher than the value reported by Wallner \textit{et al.}, respectively. The comparison has been also made using a higher threshold ($E_ {dep}^{min} = 1.5$ MeV) in the analysis of data from this work, the results being in agreement within the uncertainty with the ones obtained with the selected threshold of 250 keV.

\section{Conclusion}
The $^{238}$U(n,$\gamma$) reaction cross-section has been measured at the n\_TOF facility of CERN using an array of two deuterated-benzene scintillators. The purpose was to provide highly precise results to solve the inconsistencies among data in literature and in evaluated libraries. This cross section is in fact of extreme importance for the calculations of fast and thermal reactor parameters, and in particular the safety of nuclear reactors mainly depends on its uncertainty. 

Thanks to the low background of the experimental setup, and to the high intensity of the n\_TOF neutron flux, the cross section could be measured from 1 eV to 700 keV. The measurement has been prepared and the data analyzed with extreme accuracy and precision, in order to provide results with an uncertainty as low as possible. In particular, great efforts have been made in order to evaluate and estimate all sources of background that affected the measurement. 

As expected, resonance parameters from this work show very good agreement with data in the JEFF-3.2 and ENDF/B-VII.1 evaluated libraries. Nevertheless, for a few weak resonances significantly higher kernels were found compared to evaluated data.

For neutron energies higher than 3 keV, the cross section has been extracted from the measured yield using average parameters to determine the correction factor for self-shielding and multiple interaction. In this unresolved resonance region, the $\gamma$ background strongly affects the measurement and, since it is not possible to directly measure its contribution, it has been precisely determined by means of Monte Carlo simulations. 

For energies up to 20 keV results from this work are in fair agreement with the recent results obtained in GELINA from Kim \textit{et al.} and with the cross section recommended by Carlson \textit{et al.}, and up to 630 keV with the recent data by Ullmann \textit{et al.}. From 20 to 80 keV the cross section from this work tends to stay slightly below the evaluated data, while for higher neutron energies from 80 to 700 keV the trend is opposite, and this work yields a cross section 15 to 25\% higher than the ENDF/B-VII.1 evaluation, and 21 to 32\% higher than the JEFF-3.2 evaluation. 

A comparison with the spectrum-averaged cross-section obtained by Wallner \textit{et al.} shows good agreement for the result with the energy distribution centered at $E_n=25.3$ keV, while data from this work yield a 16\% higher value for the energy distribution peaked at $E_n=426$ keV.

\section{acknowledgment}
This work has been fully supported by the the European Commission within the FP7 project ANDES (FP7-249671). We acknowledge the Joint Research Centre in Geel of the European Commission (EC-JRC in Geel, Belgium) to have provided the high-purity sample used for the measurement. GEANT4 simulations of the neutron background have been run at the Laboratory for Advanced Computing, Faculty of Science, University of Zagreb.


\providecommand{\noopsort}[1]{}\providecommand{\singleletter}[1]{#1}%

\end{document}